# Impact Ionization and Transport Properties of Hexagonal Boron Nitride in Constant-Voltage Measurement

*Yoshiaki Hattori*[*†], *Takashi Taniguchi*[‡], *Kenji Watanabe*[‡] *and Kosuke Nagashio*[**†§]
[†]Department of Materials Engineering, The University of Tokyo, Tokyo 113-8656, Japan
[‡]National Institute of Materials Science, Ibaraki 305-0044, Japan
[§]PRESTO, Japan Science and Technology Agency (JST), Tokyo 113-8656, Japan
[*]hattori@adam.t.u-tokyo.ac.jp, [**]nagashio@material.t.u-tokyo.ac.jp

**ABSTRACT**

The electrical evaluation of the crystallinity of hexagonal boron nitride (*h*-BN) is still limited to the measurement of dielectric breakdown strength, in spite of its importance as the substrate for 2-dimensional van der Waals heterostructure devices. In this study, physical phenomena for degradation and failure in exfoliated single-crystal *h*-BN films were investigated using the constant-voltage stress test. At low electrical fields, the current gradually reduced and saturated with time, while the current increased at electrical fields higher than ~8 MV/cm and finally resulted in the catastrophic dielectric breakdown. These transient behaviors may be due to carrier trapping to the defect sites in *h*-BN because trapped carriers lower or enhance the electrical fields in *h*-BN depending on their polarities. The key finding is the current enhancement with time at the high electrical field, suggesting the accumulation of electrons generated by the impact ionization process. Therefore, a theoretical model including the electron generation rate by impact ionization process was developed. The experimental data support the expected degradation mechanism of *h*-BN. Moreover, the impact ionization coefficient was successfully extracted, which is comparable to that of $SiO_2$, even though the fundamental band gap for *h*-BN is smaller than that for $SiO_2$. Therefore, the dominant impact ionization in *h*-BN could be band-to-band excitation, not defect-assisted impact ionization.

## I. Introduction

Hexagonal boron nitride (*h*-BN) has attracted much attention as an ideal substrate for 2-dimensional (2D) van der Waals heterostructure devices with improved performance [1–4]. In addition to the high dielectric strength in *h*-BN, its high chemical stability and thermal conductivity are outstanding characteristics for electronic device applications. Therefore, electrical characterization of *h*-BN as a dielectric material is important. To date, fundamental research on dielectric constant [5, 6], tunneling current [7, 8], shot noise [9], electric field screening [10], reliability [11, 12], and breakdown strength ($E_{BD}$) [7, 8, 11, 13, 14–17] has been conducted.

Breakdown strength is the most representative property of an insulator that can be used for the evaluation of film quality and is measured easily by the time-zero dielectric breakdown test, where a voltage ramp stress is applied to the sample until catastrophic failure. $E_{BD}$ of exfoliated single-crystal *h*-BN has been reported as ~10–12 MV/cm in the out-of-plane direction [7, 8, 11, 13] with ~3 MV/cm obtained for the in-plane direction [13]. On the other hand, it has been reported that the out-of-plane $E_{BD}$ of the scalable *h*-BN grown by various growth methods is lower than that of the exfoliated sample and is limited to ~4 MV/cm [14–17] due to the presence of defects, impurities and/or grain boundaries [18, 19].

Based on the comparison with $E_{BD}$ of other materials in a general relationship between $E_{BD}$ and $\varepsilon_{BN}$ [13], $E_{BD}$ for the single-crystal *h*-BN film exfoliated from bulk crystals grown by the temperature-gradient method under a high-pressure and high-temperature atmosphere is considered to be very close to the ideal value. However, the presence of oxygen and carbon impurities with concentration of less than $10^{18}$ $cm^{-3}$ and nitrogen vacancies has been experimentally confirmed by secondary ion mass spectrometry measurements [20] and scanning tunneling microscope (STM) [21]. Despite its importance, the relationship between these defects and the insulating properties has not been elucidated as yet.

The time-dependent dielectric breakdown test, where a constant voltage or a constant current is applied to the sample until catastrophic failure, has been commonly used for the thermally oxidized silicon used as the gate insulator to allow the quality evaluation that is more sensitive than the evaluation of $E_{BD}$. This is because the key physical phenomena for degradation and failure, such as the carrier trapping to the defect states and the carrier generation due to impact ionization in $SiO_2$, can be detected though the current-time (*I*-*t*) characteristics. While still subject to some debates, based on the several models proposed so far [22–36], the degradation phenomena of $SiO_2$ have been physically formulated and the key physical properties such as trap density, and impact ionization coefficient have been evaluated quantitatively. On the other hand, almost no physical properties have been elucidated for exfoliated single-crystal *h*-BN.

In this article, ab constant-voltage stress test was performed using a high-quality single-crystal *h*-BN. The purpose of this study is to develop an appropriate theoretical model fort *h*-BN and to derive unknown physical properties, especially trap density, and impact ionization coefficient. The derived physical properties can be utilized as representative values to characterize the crystallinity of *h*-BN more sensitively than can be performed by $E_{BD}$ evaluation.

## II. Experimental method

In this study, the exfoliated single-crystal *h*-BN film was investigated by applying the electrical stress in the out-of-plane direction. **Figure 1(a)** shows the optical image of the device, where *h*-BN film with the thickness of $T_{BN}$ = 17.3 nm is sandwiched by vertical metal electrodes. Array-type electrodes enable multiple electrical measurements at different locations in the same film. The schematic of the cross-sectional device structure and the electrical measurement is illustrated in the inset of **Fig. 1(b)**. The detailed device fabrication procedure for the metal-sandwiched device has been described in reference [37]. Thin *h*-BN (10 – 30 nm) films were prepared on the poly (methyl methacrylate) (PMMA) by mechanical exfoliation technique from bulk single crystals. The targeted thin *h*-BN film was transferred on the bottom array electrodes (30-nm Au / 15-nm Cr) fabricated on 90-nm-$SiO_2$/Si wafer by electron beam (EB) lithography in advance. After the sacrificial PMMA layer was removed by acetone and isopropyl alcohol, the top electrodes were again patterned on *h*-BN by EB lithography with PMMA. Prior to metal deposition for the top electrodes (15-nm Cr / 30-nm Au), ozone treatment was performed for 5 min to remove the resist residue on the *h*-BN surface [38]. The area of the junction is typically 2-μm × 2-μm.

Alternatively, the graphite/*h*-BN/graphite device was also prepared, because an ideal stacking of 2D layered materials provides a clean interface without contaminants, such as the sacrificial polymer residue. The device was fabricated by the dry transfer technique using a PMMA/ polydimethylsiloxane (PDMS) film. The single-crystal graphite thin film was also prepared by the same method from the Kish graphite. The detail procedure of the stacking has been descried in our previous paper [39]. The stacked graphite/*h*-BN/graphite films were transferred onto quartz. Then, lead wires and pads were fabricated with Cr/Au metals by EB lithography.

The thickness of the *h*-BN film was measured using an atomic force microscope in the tapping mode. Electrical measurements were performed in vacuum (~$5.0\times10^{-3}$ Pa) at room temperature (21 – 25 °C) in the probe using semiconductor analysis. The bottom electrode was grounded in all measurements. With the exception of the experiment in **Fig. 3,** negative voltage was applied to the top electrode in all measurements. The sampling time for the constant-voltage test is 0.1 s. The voltage step and the ramping rate for the *I*-*V* measurements were 0.01 V and 0.14 V/s, respectively.

## III. Constant voltage measurements

Prior to the quantitative investigation of the time-dependence current, it should be confirmed that the current flows uniformly in the overlap area. **Figure 1(b)** shows the plot of the current density ($j$) versus electrical field ($E_{BN}$) defined by $-V_{BN}/T_{BN}$ for different electrode areas in the same film, where $V_{BN}$ is the applied voltage. All measurements match well, indicating the uniformity of current injection. Therefore, quantitative evaluation of the current is possible using the metal electrode devices. **Figure 1(c)** shows the Fowler-Nordheim (F-N) plot for the *I*-*V* character. The linear relationship indicates that the F-N tunneling current is dominant in the *I-V* test, where the slope of the line gives the barrier height ($\Phi_B$) of 2.5 ± 0.1 eV. Recently, we have found that the dominant carrier which were injected into *h*-BN by F-N tunneling in the metal/h-BN/metal structure is hole since the barrier height for the hole injection is smaller than that for the electron injection [40]. Therefore, the calculated barrier is applied for hole at the bottom Au electrode, as shown in the inset of the **Fig. 1(c)**. Moreover, the leakage current related to weak spots in *h*-BN was not measured in the detectable range before the F-N tunneling current. These results are attributed to the high crystallinity of the *h*-BN film that also enables the study of the intrinsic properties of *h*-BN.

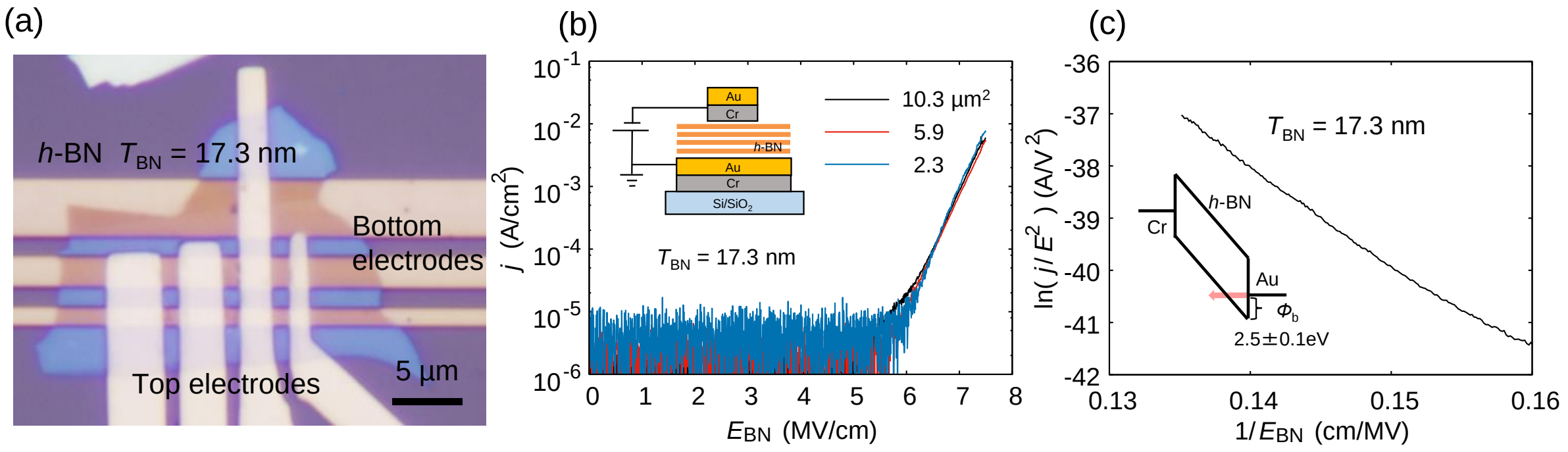

**FIG. 1** Device structure and uniformity of carrier transport. **(a)** Optical image of the typical device, where *h*-BN films are sandwiched between top and bottom electrodes. **(b)** Current density as a function of the electrical field for different electrode area. The inset of the figure is the schematic diagram for the device structure and the measurement setup. **(c)** Fowler-Nordheim plot of the *I*-*V* character. The slope of the line gives the barrier height of 2.5±0.1 eV for Au electrode.

**Figures 2(a) and (b)** show the *I-V* measurements repeated ten times and the constant-voltage stress tests, respectively. The current decreases with the iteration number of the *I-V* measurements, and this effect is especially pronounced for the first five times. The degradation of the current was also confirmed in the constant-voltage tests for 8.1 MV/cm and 8.3 MV/cm. Typically, the current decreases to ~5 % of its initial value in the constant-voltage stress test. Interestingly, the spike-like fluctuation appears in the current stronger than ~8.3 MV/cm, which may be the indication of impact ionization possess at the high field, which will be discussed in the subsequent paper. By contrast, for the electrical field stronger than ~10 MV/cm, the current increases after the slight decrease for initial 30 s, and finally leads to the catastrophic dielectric breakdown. It should be noted that *h*-BN could tolerate over 7 MV/cm stress for more than 7 hours (still without any breakdown), suggesting the high crystallinity of *h*-BN. The transition behavior in *h*-BN at high electrical stress is unique characteristic. In case of $SiO_2$, when the relatively high electrical stress is applied, the current continually decreases and the breakdown suddenly occurs, where the time for breakdown depends on the applied stress [23].

Next, the origin of the current degradation is investigated. If the holes are trapped in *h*-BN, the trapped holes will lower the electrical field near the anode, resulting in the decrease of the current. On the other hand, the situation is reversed for the case of electron trapping, as schematically shown in **Fig. 3(a)**. Therefore, removal of the accumulated carrier was attempted by thermal annealing. **Figure 3(b)** shows the effect of annealing at 100 °C for 2 hours in the Ar/$H_2$ forming gas. The *I-V* character degraded by the application of the electrical stress of 7.0 MV/cm for 1,800 s was recovered by the subsequent annealing due to thermal detrapping. Note that the thermal annealing tests were conducted in the graphite/*h*-BN/graphite device using relatively thick graphite (~30 nm) as electrodes, as shown in the inset of **Fig. 3(b)**, because the oxidation of Cr during the annealing was confirmed in the electrical test. Regardless of electrode materials, similar current reductions are observed. Based on these experiments, it is concluded that the current reduction is due to the hole trapping and not due to the formation of permanent defects in *h*-BN. This can also be supported by the fact that the decreased current shown in **Fig. 2** recovers ~90% of the initial value by simply leaving the sample at the atmospheric pressure for several days.

The location of the trapped holes was investigated according to the previous study of $SiO_2$ [22]. **Figure 3(c)** shows the *I-V* behavior before and after the application of the 6.6-MV/cm stress for 900 s. The clear difference in the voltage shift for the positive voltage ($\Delta V^+$) and negative voltage ($\Delta V^-$) was observed, providing evidence for the deviation of the centroid of holes. It should be noted that both polarities of voltage were measured individually using the thermal detrapping technique to minimize the hole trapping during the *I-V* test. The location of the centroid of holes is given by [22]

$$x_p = \frac{\Delta V^+}{\Delta V^- + \Delta V^+} T_{BN}, \quad (1)$$

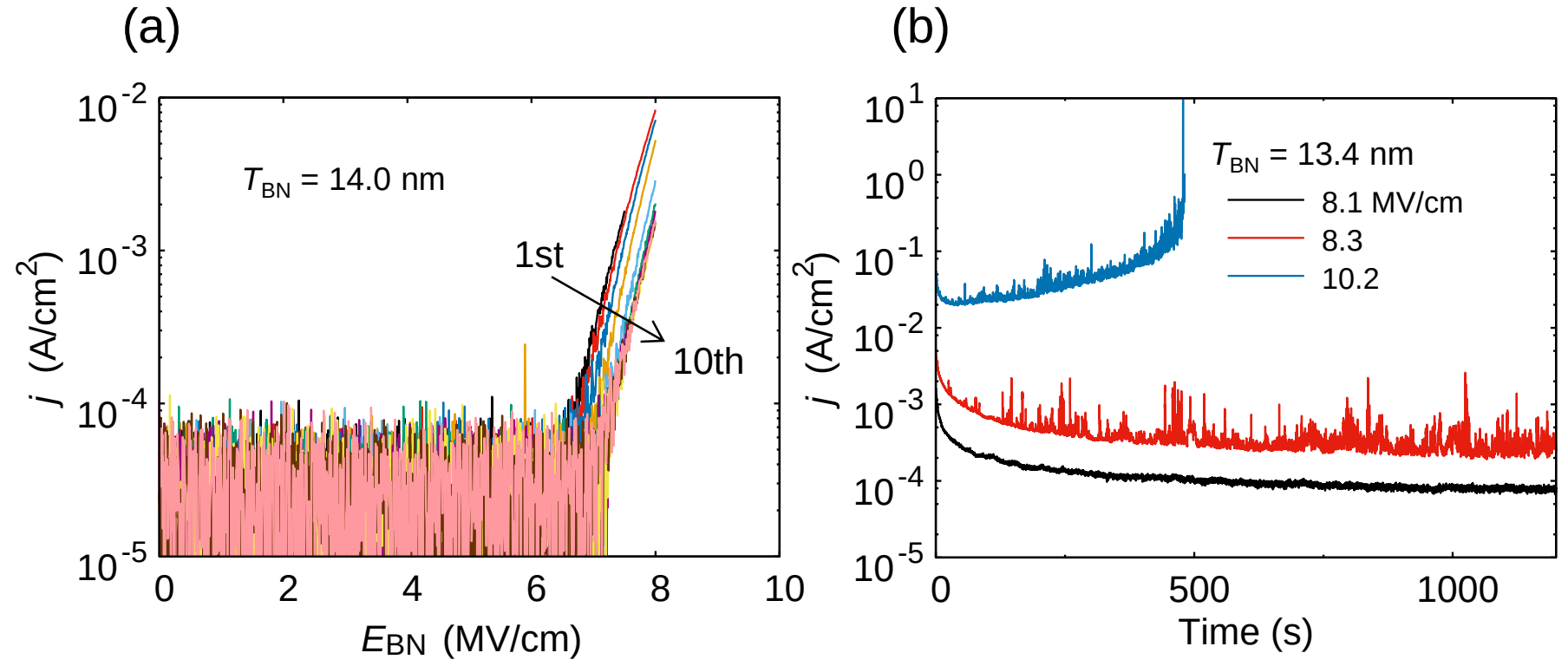

**FIG. 2** Repeated *I-V* measurement and the constant-voltage test. **(a)** *I-V* measurement repeated ten times. The figure is converted to current density and electrical field. The current decreases gradually with the number of measurements. **(b)** Constant voltage test at different electrical fields.

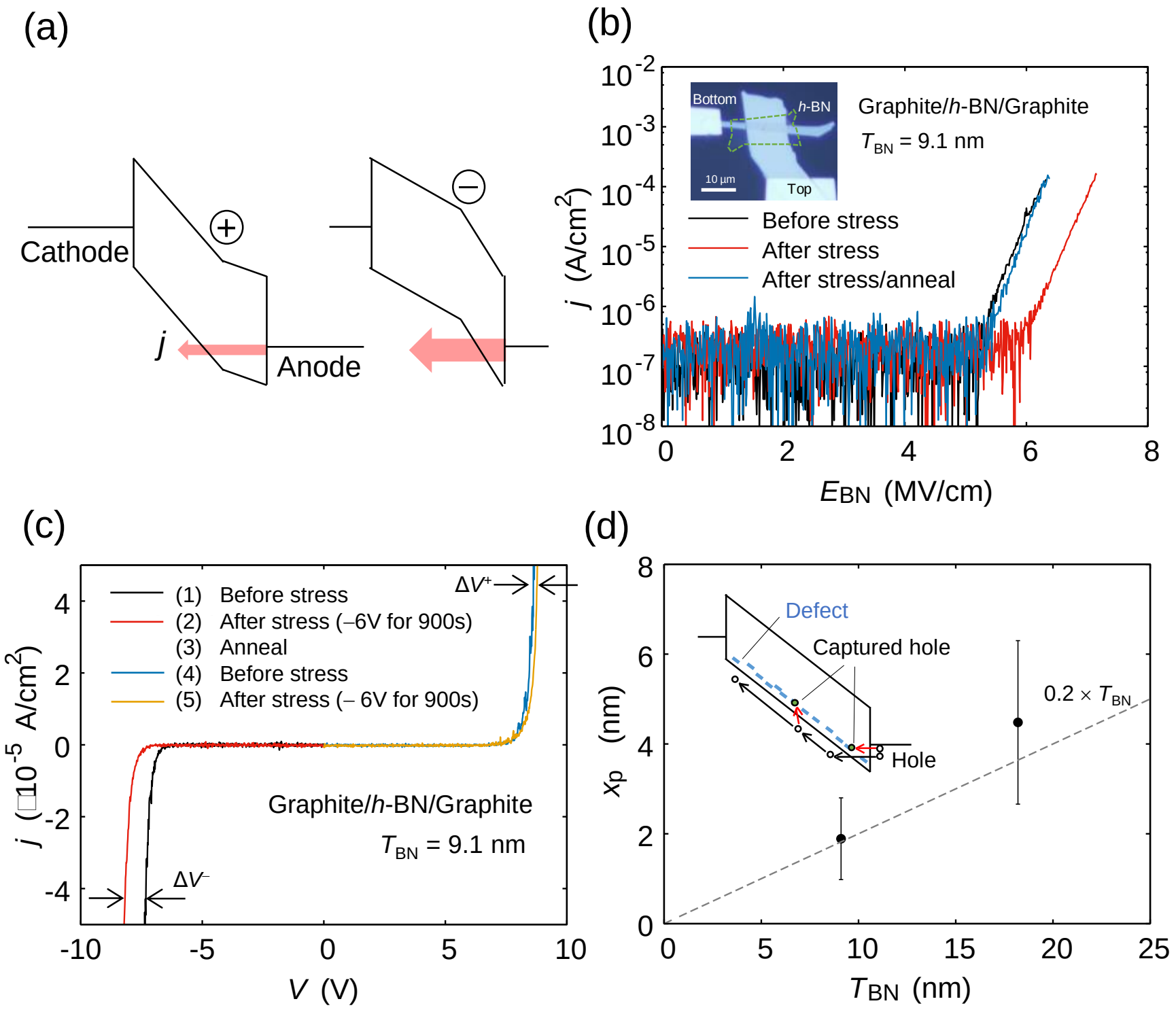

**FIG. 3** Centroid location of the trapped hole. **(a)** Schematic of local electrical fields for *h*-BN with hole and electron traps. **(b)** *I*-*V* measurements before and after applying the stress of 6.6 MV/cm for 1800 s. **(c)** *I*-*V* character for both polarities in linear scale before and after the stress. First, *I*-*V* in negative polarity was measured before and after the negative stress of −6 V (6.6 MV/cm) for 900 s (holes are injected from the bottom electrode). Subsequently, the sample was annealed at 100 °C for 2 hours for recovery. Then, *I*-*V* in positive polarity was measured in the same manner. **(d)** Centroid location of hole as a function of the *h*-BN thickness. Note that all the experiments in Fig. 3 were performed using graphite/*h*-BN/graphite device shown in the inset of **(b)**.

where $x_p$ is the distance from the anode to the centroid of holes. Figure 3(d) shows $x_p$ for the two samples with different thickness, indicating that $x_p$ is found to be close to the anode. The detailed experimental procedure is described in Supplemental Material Fig. S1. In addition, the thicker sample has larger $x_p$, which relation is independent of the stress time and the polarity of the stress (Supplemental Material Fig. S1). The hole trapping might occur in the tunneling or the transportation in *h*-BN, as schematically illustrated in the inset of Fig. 3(d). Since the number of the trapped hole in thick *h*-BN should be larger than that in thin *h*-BN, $x_p$ should increases with increase of the $T_{BN}$ as an average value.

Let us now return to the discussion of the current degradation shown in **Fig. 2(b)**. The current initially decreases and gradually saturates with time, which means that a maximum density of the neutral trap sites for hole ($N_t$) exists. Here, it was investigated whether $N_t$ depends on the electrical field. The schematic of the applied voltage as a function of time is provided in **Fig. 4(a)**. The applied voltage is increased stepwise by 0.2 V from 7.0 MV/cm until the breakdown at the 900-s step using a bias hold function in the semiconductor device analyzer. In the beginning of the test, the current reduction was clearly observed as shown in the figure for 7.0-MV/cm stress. However, current reduction was not observed for electrical fields more than ~8 MV/cm. This result indicates that all trap sites for holes have been already filled prior to the application of the 8-MV/cm stress and $N_t$ is therefore independent of the electrical field. If $N_t$ increases with increasing electrical filed, the current should decrease with time at the beginning of each step for the elevated electrical fields. It should be noted that the present experiment was conducted continuously for only one location of the array-type electrodes for *h*-BN, unlike in **Fig. 2(b)**.

Finally, the current at the high electrical field is focused. The currents saturated at different electrical fields as shown by open circles in **Fig. 4(a)**, were plotted as a function of the electrical field in **Fig. 4(b)** and were compared to the calculation using the standard F-N tunneling model. Interestingly, the measured current exceeds the standard F-N current at ~8.5 MV/cm. The origin of this current enhancement is explained as follows. At low electrical fields, holes trapped in *h*-BN reduce the electrical field near the anode, resulting in the current reduction. As the electrical field is increased, the electron-hole pair will be generated due to the impact ionization by hot holes. The generated holes will drift to the cathode because the neutral trap for hole sites have already been occupied, while the generated electron will be trapped in the neutral trap. This enhances the electrical field near the anode, as schematically shown in **Fig. 3(a)**, and results in the current increase. Finally, catastrophic dielectric breakdown will occur. As discussed above, the current enhancement suggests the occurrence of the impact ionization. However, it is difficult to evaluate the impact ionization coefficient directly from the experimental data. Therefore, to further support the above expectation and understand the physical mechanism of degradation of *h*-BN in more detail, a theoretical model was developed for quantitative investigations described in the next section.

## IV. Theoretical model

As discussed in the previous section, the trapped carriers enhance or lower the field in the *h*-BN. For simplicity, the charge sheet model is used here [22–34]. The band diagram for the present model is shown in **Fig. 5(a)**. When an electrical field is applied across *h*-BN, the internal field is expressed piecewise by Gauss's law using the following equations [23, 24].

$$E_{\mathrm{a}} = E_{\mathrm{BN}} - \frac{qp_{\mathrm{t}}}{\varepsilon_{\mathrm{BN}}}\left(1 - \frac{x_{\mathrm{p}}}{x_{\mathrm{BN}}}\right) + \frac{qn_{\mathrm{t}}}{\varepsilon_{\mathrm{BN}}}\left(1 - \frac{x_{\mathrm{n}}}{x_{\mathrm{BN}}}\right), \quad (2)$$

$$E_{\mathrm{m}} = E_{\mathrm{a}} - \frac{qn_{\mathrm{t}}}{\varepsilon_{\mathrm{BN}}}, \quad (3)$$

$$E_{\mathrm{c}} = E_{\mathrm{a}} + \frac{qp_{\mathrm{t}}}{\varepsilon_{\mathrm{BN}}}, \quad (4)$$

where $p_{\mathrm{t}}$ and $n_{\mathrm{t}}$ are the densities of trapped holes and electrons, respectively. $x_{\mathrm{n}}$ represents the centroid of electrons. $\varepsilon_{\mathrm{BN}}$ is the permittivity of *h*-BN. The subscripts $a$, $m$ and $c$ indicate the anode, middle and cathode, respectively. The F-N tunneling current ($j$) is regarded as the dominant current between the electrodes. Because the tunneling region near the anode is generally not triangular, the F-N tunneling current for arbitrarily shaped potential barriers should be used [23, 24, 41–43]. This is given by

$$j = \frac{C_1}{B^2}\exp(-C_2 A), \quad (5)$$

where $A$ and $B$ are calculated from the shape of the potential barrier $\Phi(x)$ based on the Fermi energy of the anode; $C_1$ and $C_2$ are the material constants that depend only on the hole effective mass ($m^*$) in *h*-BN. $m^*$ is set to be $0.5\times m_0$ in the calculation [24], where $m_0$ is the electron mass in vacuum. A detailed description of Eq. (5) is provided in Supplemental Material **Note 1**. The image-force lowering effect is ignored because this effect is negligible for the electrical field magnitudes considered in this paper [44–47]. Eqs. (2–5) indicate that trapped holes lower the $E_{\mathrm{a}}$, which in turn lowers the current, resulting in negative feedback. Similarly, it is understood that the accumulation of electrons gives rise to positive feedback for dielectric breakdown.

The rate equation for holes captured by neutral traps is given by [24, 25],

$$\frac{dp_t}{dt} = \frac{\sigma_t j}{q}(N_t - p_t)\ , \quad (6)$$

where $\sigma_{\mathrm{t}}$ is the trap capture cross-section area for the hole; and $t$ is time. The detrapping of the holes is not considered here. Electrons are generated by the impact ionization process that is activated at the higher field. The generated total

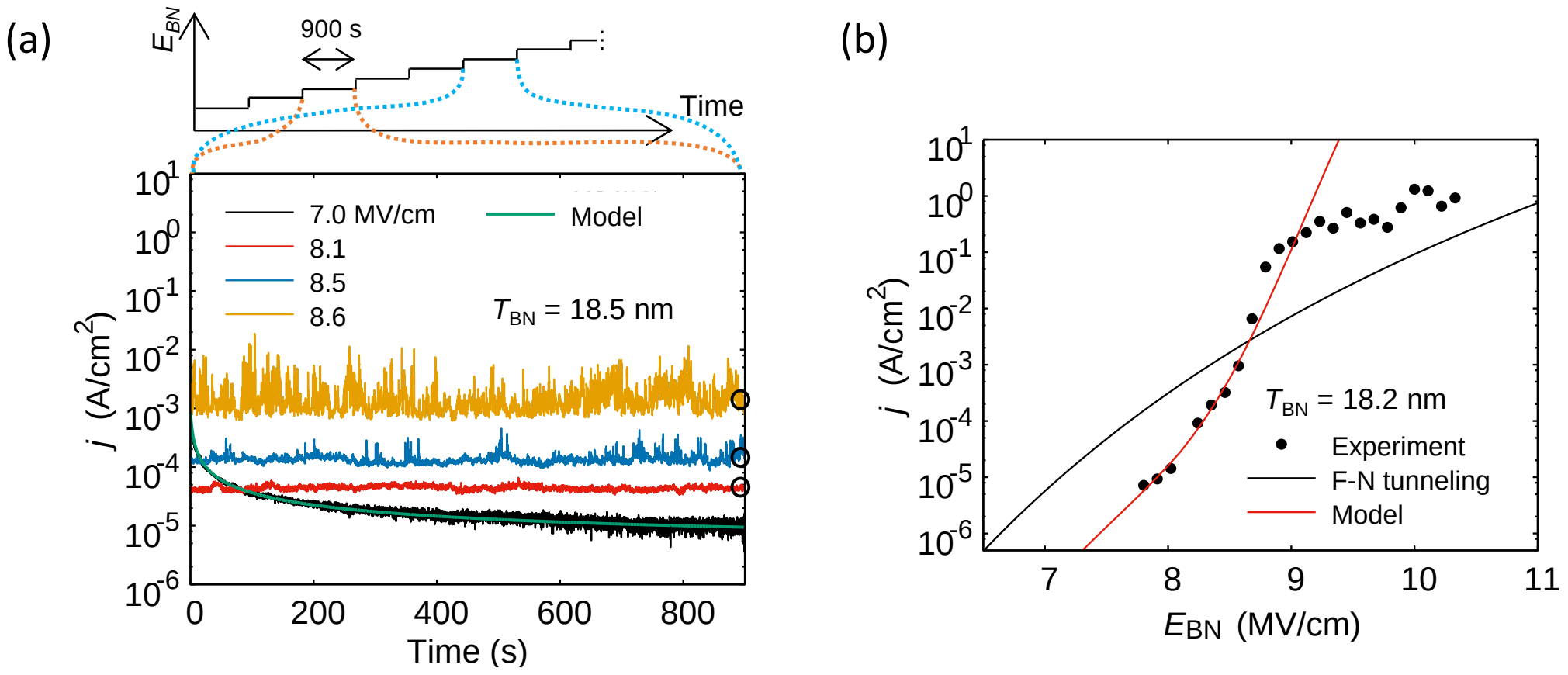

**FIG. 4** Current density in the equilibrium state depends on field. **(a)** Time-dependent current under the stair-like constant-voltage stress. The applied voltage is continuously increased every 900-s from 7.0 MV/cm until breakdown. The green solid line indicates the simulation by the electron trapping model. Open circles indicate the current densities at the steady state for different electrical fields. **(b)** Current density at the steady state as a function of electrical fields with the simulation by the standard F-N tunneling current model.

amount of electrons ($n_G$) is written as [23, 24]

$$n_G = \frac{1}{q}\int_0^t j\left(\int_{x_t}^{x_{BN}} \alpha dx\right) dt, \tag{7}$$

where $\alpha$ is the impact ionization coefficient that depends on the field $E(x)$. $x_t$ represents the tunneling distance, which is calculated geometrically with the local field (Supplemental Material **Note 1**). The number of trapped electrons in $h$-BN changes with the generation rate, recombination rate, and electron drift velocity according to a rate equation [35]. In the present model, only generation and recombination processes are considered [26–28, 36]. The assumption corresponds to the so-called impact ionization-recombination (IR) model to explain the breakdown for insulators, by which the reasonable agreement with the experimental data has been reported [28]. The recombination process occurs when holes drifting in the valence band of $h$-BN are captured by the electrons. The rate is given by [24, 29, 30],

$$\frac{dp_R}{dt} = \frac{\sigma_R j}{q}(n_G - p_R), \tag{8}$$

where $p_R$ is the density of the total holes trapped by the electrons starting at $t$ = 0 and $\sigma_R$ is the trap capture cross section area for recombination. The density of trapped electrons ($n_t$) at certain time in $h$-BN is given by,

$$n_t = n_G - n_R, \tag{9}$$

where the electron recombination rate ($n_R$) is equal to $p_R$. Note that Eqs. (8) and (9) are valid only if the number of neutral trap sites for electrons far exceeds the number of occupied sites [29].

These combined nonlinear Eqs. (2)–(9) are solved by the fourth-order-Runge-Kutta method for the comparison with the experiment data. The initial conditions are $p_t$ = 0 and $n_t$ = 0. Namely, the $h$-BN film is assumed to be free of charge in the initial state. Although most of the physical parameters of $h$-BN used for the calculation have not been reported yet, the physical parameters including the impact ionization coefficient are estimated from the experimental data step by step in the next section. Usually, the capacitance-voltage measurement [27, 29, 33] or a high-energy source such as γ-ray [30] is required to estimate the impact ionization coefficient for an insulating film. However, the developed analysis enables us to obtain them only from *I-t* measurement, because the unique *I-t* characteristic in $h$-BN, that is, the steady-state current at high electrical field, includes the information on the impact ionization process. The physical parameters can be extracted by arranging the above equations and fitting the experimental data.

## V. Comparison with the experiments

First, let us consider the case of low electrical field stress. Assuming that the impact ionization does not occur

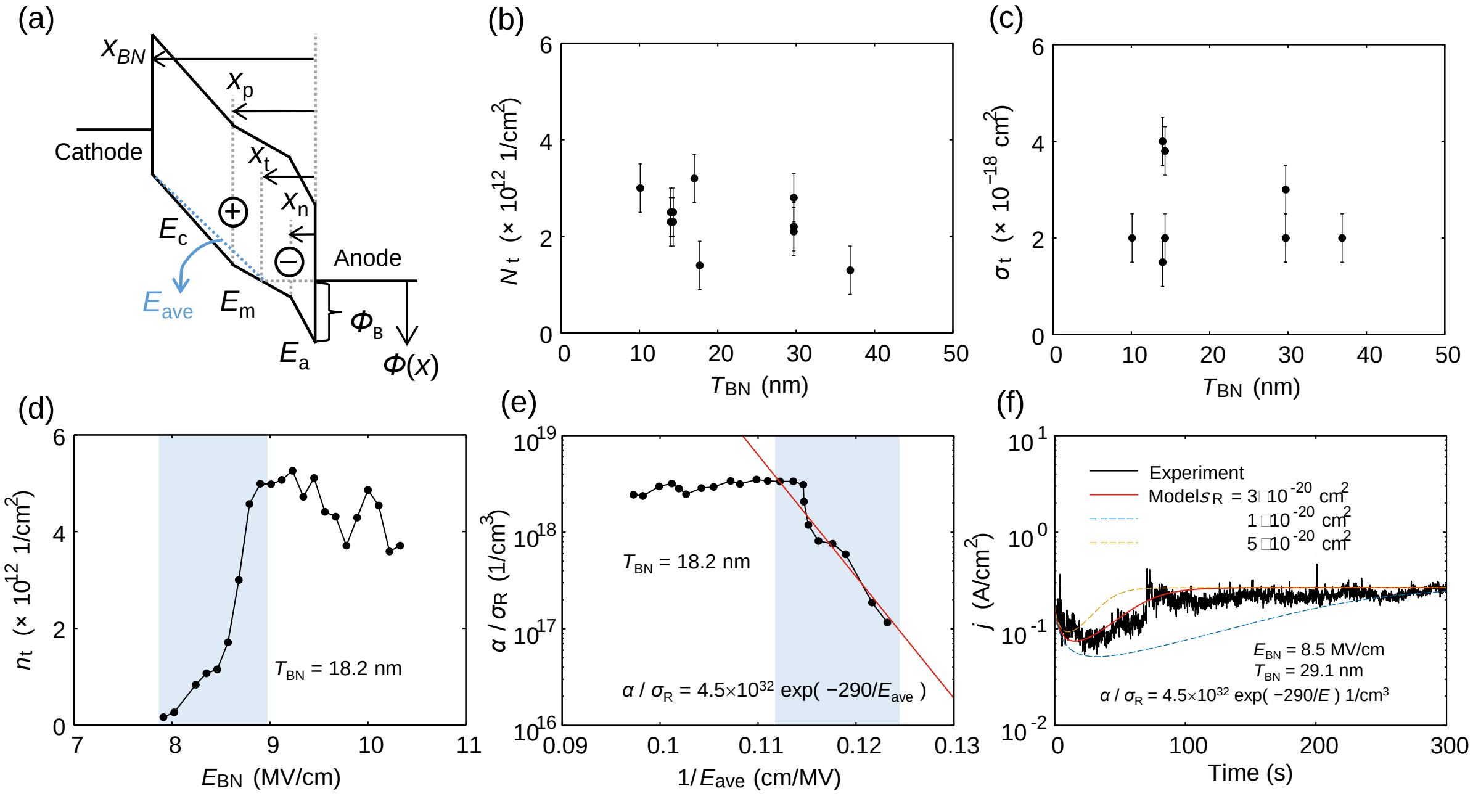

**FIG. 5** Analysis of experimental data using the theoretical model. **(a)** Energy band diagram for $h$-BN with trapped hole and electron under the electrical field. **(b, c)** Density of total neutral trap sites and capture cross section area for hole as a function of $h$-BN thickness. **(d)** Density of trapped electron at the steady state at different electrical fields. **(e)** $\alpha/\sigma_R$ as a function of $1/E_{ave}$. The value of $\alpha/\sigma_R$ corresponds to amplitude of the impact ionization process. The slop in the enhancement range for electron accumulation characterizes impact ionization coefficient of $h$-BN. **(f)** Constant-voltage test for high electrical field to separate $\alpha_0$ and $\sigma_R$.

he

estimation of two independent fitting parameters $N_t$, and $\sigma_t$ by reproducing the current degradation behavior obtained at the constant-voltage stress. An example of this fitting for the electric field of 7 MV/cm is shown in **Fig. 4(a)**, where $\varepsilon_{BN}$ is set to 3.38 [48] and $x_p$ is set to $0.2\times T_{BN}$ in accordance with the experimental results in **Fig. 3(d) [22]**. The calculation results are in good agreement with the experimental data. Since $N_t$ and $\sigma_t$ characterize different physical properties, that is, the amount of current degradation and the current degradation rate, respectively, the accuracy for estimated $N_t$ and $\sigma_t$ is high enough to quantitatively discuss the physical mechanism of degradation of *h*-BN using these two physical properties, as shown in Supplemental Material **Fig. S2**. Moreover, the dependence of the estimated $N_t$ and $\sigma_t$ on $T_{BN}$ is shown in **Figs. 5(b) and (c)**, respectively. Both physical properties are independent of $T_{BN}$ in the measured range.

Next, the case of a high electrical field is considered. To obtain $\sigma_R$ and $\alpha$, the governing equations should be simplified. The important point realized from the experiments is that the current saturates with time, that is, the steady state condition is reached where the electron trapping and detrapping rates are equal. First, $n_t$ is calculated as a function of $E_{BN}$ using Eqs. (2)–(5). The difference ($\Delta j$) between the measured steady-state current and the theoretical F-N tunneling current is attributed to the accumulation of electrons and holes in *h*-BN. In addition, $p_t$ is set to be equivalent with $N_t$ at the steady state condition where the neutral trap sites for holes are fully occupied. Therefore, if $x_n$ is known, $n_t$ is uniquely determined as an inverse problem. For $x_n$ of $SiO_2$, it has been reported to that the centroid of accumulated carrier which increases the current is located near the electrode for F-N tunneling injection side [29–32]. Therefore, $x_n$ was assumed to be 2 nm in this simulation. It is noted that $x_n$ is not sensitive to the later calculations (Supplemental Material **Fig. S3**). **Figure 5(d)** shows $n_t$ as a function of $E_{BN}$ calculated by the dichotomy method using the experimental data plotted in **Fig. 4(b)**. $n_t$ increases rapidly at 8 MV/cm and saturates at ~9 MV/cm. The saturation may be explained as due to all neutral trap sites for electrons being occupied. Therefore, the density of the neutral trap sites for electrons is considered to be ~$4.0\times10^{12}$ cm$^{-2}$, which is comparable to $N_t$.

Next, the governing equations are modified for the steady state condition. Because $dn_t/dt$ becomes zero, the following relation is obtained by differentiating Eq. (9).

$$\frac{dn_G}{dt} = \frac{dn_R}{dt}. \tag{10}$$

By calculating Eqs. (8) – (10), the following equation can be obtained.

$$\frac{j}{q}\left(\int_{x_t}^{x_{BN}} \alpha(x,E)dx\right) = \frac{\sigma_R j}{q} n_t. \tag{11}$$

(a)

| Physical parameters | Typical obtained values |
|---|---|
| $x_p$ | $0.2 \times T_{BN}$ |
| $N_t$ | $2.0\pm1.0\times10^{12}$ (1/cm$^2$) |
| $\sigma_t$ | $2.5\pm1.5\times10^{-18}$ (cm$^2$) |
| $\sigma_R$ | $3.0\pm2.0\times10^{-20}$ (cm$^2$) |
| $\alpha(E)$ | $\alpha(E) = \alpha_0 \exp(-H/E)$<br>$\alpha_0 = 1.4\times10^{13}$ (1/cm)<br>$H = 2.9\times10^8$ (V/cm) |

(b)

| Materials | Bandgap (eV) | $E_{BD}$ (MV/cm) |
|---|---|---|
| Ge | 0.7 [61] | 0.12 [65] |
| Si | 1.1 [61] | 0.3 [61] |
| InP | 1.3 [61] | 0.45 [61] |
| GaAs | 1.4 [61] | 0.4 [66] |
| 4H-SiC | 3.2 [62] | 3.3(//c) [67]<br>2.5(⊥c) [67] |
| GaN | 3.4 [62] | 2.7 [68] |
| Diamond | 5.5 [62] | 5 – 10 [61, 62] |
| *h*-BN | 6 [63] | 12(//c) [13]<br>3(⊥c) [13] |
| $SiO_2$ | 9 [64] | 12 [70] |

(c)

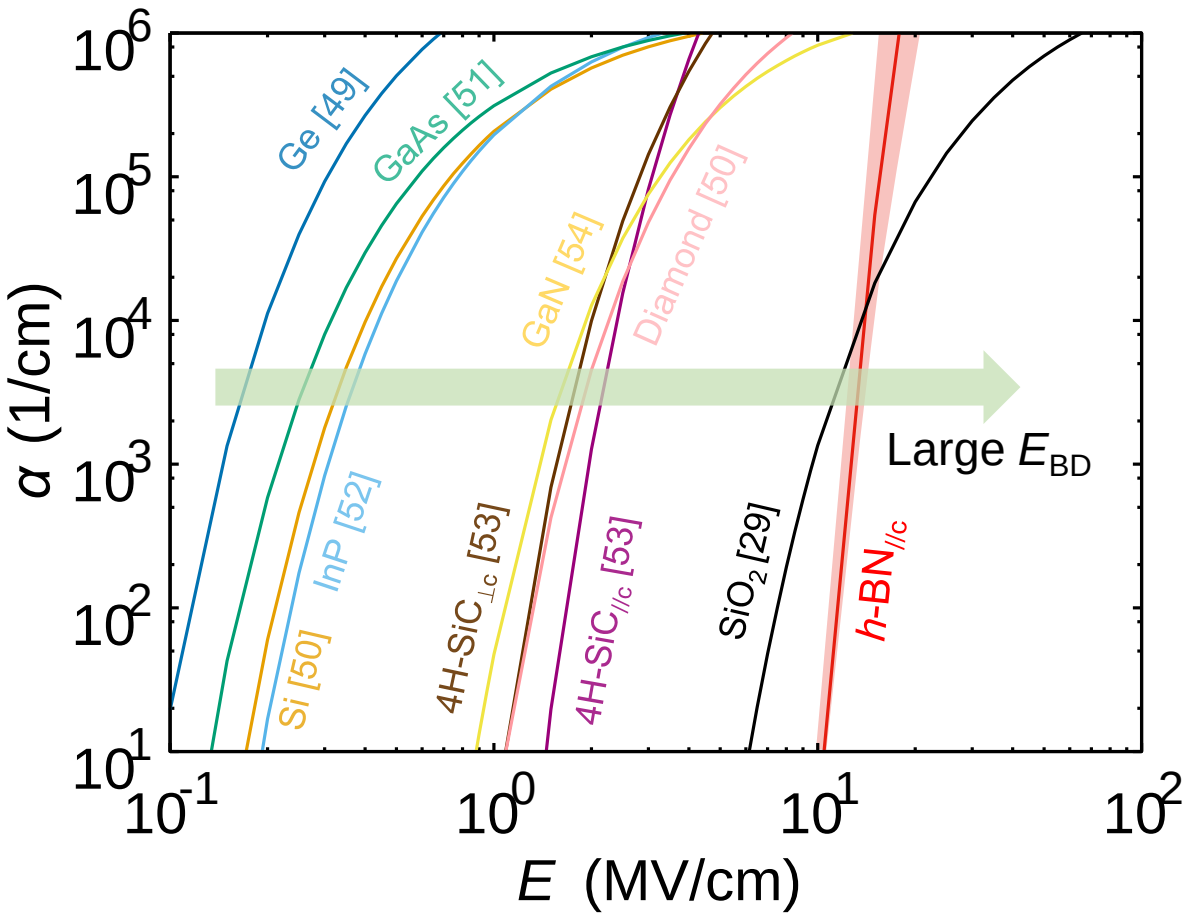

**FIG. 6** Comparison of impact ionization coefficient. **(a)** Physical parameters obtained in the present simulation. **(b)** Band gap and dielectric breakdown field for various materials. **(c)** Impact ionization coefficient as a function of electrical fields for various materials, where the all values are for hole α except for $SiO_2$. In terms of the errors for the estimated physical parameters, the error in α is shown by the hatched region in c, while the errors of $N_t$ and $\sigma_t$ are obtained from the deviation for different samples.

For simplification, the electrical field from $x_t$ to $x_{BN}$ is assumed to be spatially constant and is denoted as $E_{ave}$, defined by the following equation,

$$E_{ave} = -\frac{\phi(x_{BN})-\phi(x_t)}{x_{BN}-x_t}. \tag{12}$$

Although $E_{ave}$ is calculated analytically (Supplemental Material **Note 1**), $E_{ave}$ is practically nearly equal to $E_{BN}$. Then, Eq. (11) becomes,

$$\frac{\alpha(E_{ave})}{\sigma_R} = \frac{n_t}{x_{BN}-x_t}. \tag{13}$$

This relationship indicates the character of $\alpha(E_{ave})$, if $\sigma_R$ is independent of the electrical field. Here, it is well known that $\alpha(E)$ is empirically expressed by the following standard form,

$$\alpha(E) = \alpha_0 \exp(-\frac{H}{E}), \tag{14}$$

where the prefactor $\alpha_0$ and the enhancement factor $H$ are constants [29, 49–55]. The slope of the logarithm of $\alpha(E)$ versus $1/E$ provides the enhancement factor. **Figure 5(e)** shows the plot of $\alpha(E_{ave})/\ \sigma_R$ versus $1/E_{ave}$. $H$ and the $\alpha_0/\sigma_R$ are calculated to be 290 MV/cm and $4.5\times10^{32}$ 1/cm$^3$ from the slope and the intercept of the slop in the hatched range, respectively. The separation of $\alpha_0$ and $\sigma_R$ is possible by fitting experimental data at the non-steady state condition at high electric field. **Figure 5(f)** shows the time-dependent current density at 8.5 MV/cm obtained by simulation. By selecting $\sigma_R = 3\times10^{-20}$ cm$^2$, $\alpha_0$ is determined to be $1.4\times10^{13}$ cm$^{-1}$.

The physical parameters obtained in this simulation are summarized in **Fig. 6(a)**. Finally, numerical simulation for **Fig. 4(b)** was conducted using the obtained physical parameters. The calculations could reproduce the experiment in the range of less than 9 MV/cm, where Eqs. (8) and (9) are valid. Therefore, the proposed theoretical model strongly supports the physical mechanism for the degradation of $h$-BN for both low and high electrical fields assumed in the last part of the experimental section. It should be emphasized that based on the experimental results, the present model assumes that no permanent defects are formed under the electrical field and the carriers are trapped on the initial defects without charge introduced during the $h$-BN growth. Therefore, let us discuss the defect sites that exist in the present single-crystal $h$-BN film. Because the holes trapped in $h$-BN were detrapped by long-term retention at the room temperature or the thermal annealing at 100 °C, the trap sites for holes are considered to be shallow traps. So far, the presence of nitrogen vacancies and carbon impurities has been confirmed experimentally [20, 56, 57]. In addition to the theoretical studies, the defect levels for nitrogen vacancies ($V_N$) are expected to be less than ~1.0 eV from the bottom of conduction band [58–60]. On the other hand, the defect level for substitutional carbon impurities on the nitrogen site ($C_N$) is deep, that is, ~1.0 eV from the top of valence band [59]. Therefore, it is speculated that the trap sites for the electrons in the study are ascribed to nitrogen vacancies. On the other hand, the trap sites for holes may be ascribed to the carbon impurity since the carbon impurity is one of main impurities even for the single-crystal $h$-BN of the present quality. The trap density has been reported to be in the range of ~$10^9$ – $10^{10}$ cm$^{-2}$ by STM with the help of graphene conductive cover layer, which is smaller than the estimate in this study (~$10^{12}$ cm$^{-2}$). Although further investigations are required for the estimation of trap site densities and the relationship between the trap sites and the type of defects, from the viewpoint of the suppression for the dielectric breakdown, the reduction of trap sites for electrons is critical because electron trapping enhances the electrical field in $h$-BN.

## VI. Impact ionization coefficient

The impact ionization coefficient is fundamental intrinsic parameter for a defect-free ideal material. **Figure 6(c)** shows the comparison of impact ionization coefficients in variety of materials including semiconductors [29, 49–55], where all values are obtained for the impact ionization coefficient for hole except for $SiO_2$. The physical parameters required to calculate the impact ionization coefficient for other materials are also listed in Supplemental Material Fig. S4. Generally, materials with higher band gap have higher $E_{BD}$, as shown in **Fig. 6(b)**, which is explained simply by band-to-band impact ionization. **Figure 6(c)** indicates that it is easy to cause the impact ionization by smaller electrical field for materials with smaller band gap. The direct origin for the positive feedback to the dielectric breakdown is the enhancement of the electrical field due to the carrier trapping. Therefore, it is considered that materials with smaller band gap have higher possibility to cause the dielectric breakdown because the generation of carriers which enhance the electrical field are easily achieved by impact ionization even at the lower electrical field.

Here, let us compare $h$-BN to $SiO_2$. Experimentally, α and $E_{BD}$ of $h$-BN for the $E//c$ direction are comparable to those of $SiO_2$, even though the fundamental band gap for $h$-BN is smaller than that for $SiO_2$. Therefore, it is suggested that the dominant impact ionization in $h$-BN is band-to-band excitation, not defect-assisted impact ionization. Next, let us compare $h$-BN to 4H-SiC in terms of the crystal anisotropy. Anisotropy in α and $E_{BD}$ for 4H-SiC has been reported as shown in **Fig. 6(b) and (c)**. Since larger anisotropy in $E_{BD}$ has been reported for $h$-BN [13], it is strongly expected that $h$-BN also has larger anisotropy in α.

## VII. Conclusions

Based on both experiments and theoretical simulation, physical phenomena underlying the degradation and failure in exfoliated single-crystal *h*-BN films are explained as follows. At low electrical field, holes are injected to the valence band of *h*-BN by Fowler-Nordheim tunneling and are soon trapped to the shallow trap sites. The trapped holes reduce the electrical field near the cathode, resulting in the reduction of the current. As the electric field is increased, the electron-hole pair is generated due to the impact ionization by hot holes. The generated holes drift to the cathode, while the generated electrons are trapped in the defect sites near the conduction band in *h*-BN. This enhances the electrical field near the anode, resulting in an increased current. Finally, catastrophic dielectric breakdown occurs. Therefore, from the viewpoint of the suppression of the dielectric breakdown, the reduction of trap sites for electrons is critical because electron trapping enhances the electrical field in *h*-BN.

Moreover, the successful development of the theoretical model enables us to extract the physical properties such as impact ionization coefficient, total trap density, trap capture cross section, and recombination. These derived physical properties can be utilized as representative values to characterize the crystallinity of *h*-BN more sensitively than the current measurements relying on $E_{BD}$ evaluation.

## ACKNOWLEDGMENTS

This research was supported by the JSPS Core-to-Core Program, A. Advanced Research Networks, JSPS KAKENHI Grant Numbers JP17K14656, JP25107004, JP26886003, JP16H04343, JP16K14446, JST PRESTO Grant Number JPMJPR1425, Mikiya Science and Technology Foundation, and Mizuho Foundation for the Promotion of Sciences.

## AUTHOR INFORMATION

Corresponding Authors

Email: *hattori@adam.t.u-tokyo.ac.jp, **nagashio@material.t.u-tokyo.ac.jp

# Supplemental Material

## Impact Ionization and Transport Properties of Hexagonal Boron Nitride in Constant-Voltage Measurement

*Yoshiaki Hattori*[*†], *Takashi Taniguchi*[‡], *Kenji Watanabe*[‡] *and Kosuke Nagashio*[**†§]

[†]Department of Materials Engineering, The University of Tokyo, Tokyo 113-8656, Japan

[‡]National Institute of Materials Science, Ibaraki 305-0044, Japan

[§]PRESTO, Japan Science and Technology Agency (JST), Tokyo 113-8656, Japan

[*]hattori@adam.t.u-tokyo.ac.jp, [**]nagashio@material.t.u-tokyo.ac.jp

**Supplemental Material Note 1: Detail description of Fowler-Nordheim tunneling current.**

The F-N tunneling current for the potential barrier with arbitrary shapes is given as follows [23]:

$$j_{\mathrm{FN}} = \frac{C_1}{B^2}\exp(-C_2 A), \qquad \text{(S1)}$$

where $C_1$ and $C_2$ are constants given by

$$C_1 = \frac{qm}{2\pi h m^*}, \qquad \text{(S2)}$$

$$C_2 = \frac{4\pi\sqrt{2m^*}}{h}, \qquad \text{(S3)}$$

while $A$ and $B$ are given as the following forms, calculated from the shape of the potential barrier $\Phi(x)$ based on the Fermi energy of cathode.

$$A = \int_0^{x_t} \sqrt{q\Phi(x)}\, dx, \tag{S4}$$

$$B = \int_0^{x_\mathrm{t}} \frac{1}{\sqrt{q\Phi(x)}} dx, \tag{S5}$$

where $x_\mathrm{t}$ is the tunneling distance. $\Phi(x)$ is given piecewise as follows,

(Area 1) For $x < 0$,

$$\Phi(x) = 0. \tag{S6}$$

(Area 2) For $0 \leq x < x_\mathrm{n}$,

$$\Phi(x) = \Phi_\mathrm{B} - E_\mathrm{a} x. \tag{S7}$$

(Area 3) For $x_\mathrm{n} \leq x < x_\mathrm{p}$,

$$\Phi(x) = \Phi_\mathrm{B} - E_\mathrm{a} x_\mathrm{n} - E_\mathrm{m}(x - x_\mathrm{n}). \tag{S8}$$

(Area 4) For $x_\mathrm{p} \leq x \leq x_\mathrm{BN}$,

$$\Phi(x) = \Phi_\mathrm{B} - E_\mathrm{m} x_\mathrm{n} - E_\mathrm{m}\left(x_\mathrm{p} - x_\mathrm{n}\right) - E_\mathrm{c}\left(x - x_\mathrm{p}\right). \tag{S9}$$

(Area 5) For $x < x_\mathrm{BN}$,

$$\Phi(x) = -E_\mathrm{a} x_\mathrm{n} - E_\mathrm{m}\left(x_\mathrm{p} - x_\mathrm{n}\right) - E_\mathrm{c}\left(x_\mathrm{BN} - x_\mathrm{p}\right). \tag{S10}$$

*A, B* and, $x_\mathrm{t}$ are given separately by four kinds of situations according to $\Phi(x)$ as follows,

(Case 1) For $\Phi(x_\mathrm{n}) \leq 0$,

$$x_\mathrm{t} = \frac{\Phi(x_\mathrm{n})}{E_\mathrm{a}}, \tag{S11}$$

$$A = \frac{2}{3} q^{\frac{1}{2}} \left( \Phi_\mathrm{B}^{\frac{3}{2}} \frac{1}{E_\mathrm{a}} \right), \tag{S12}$$

$$B = 2q^{-\frac{1}{2}} \left( \Phi_\mathrm{B}^{\frac{1}{2}} \frac{1}{E_\mathrm{a}} \right). \tag{S13}$$

(Case 2) For $\Phi\left(x_\mathrm{p}\right) \leq 0 < \Phi(x_\mathrm{n})$,

$$x_\mathrm{t} = \frac{\Phi(x_\mathrm{n})}{E_\mathrm{m}} + x_\mathrm{n}, \tag{S14}$$

$$A = \frac{2}{3} q^{\frac{1}{2}} \left\{ \Phi_{\mathrm{B}}^{\frac{3}{2}} \frac{1}{E_{\mathrm{a}}} - \Phi(x_{\mathrm{n}})^{\frac{3}{2}} \left( \frac{1}{E_{\mathrm{a}}} - \frac{1}{E_{\mathrm{m}}} \right) \right\}, \tag{S15}$$

$$B = 2q^{-\frac{1}{2}} \left\{ \Phi_{\mathrm{B}}^{\frac{1}{2}} \frac{1}{E_{\mathrm{a}}} - \Phi(x_{\mathrm{n}})^{\frac{1}{2}} \left( \frac{1}{E_{\mathrm{a}}} - \frac{1}{E_{\mathrm{m}}} \right) \right\}. \tag{S16}$$

(Case 3) For $\Phi(x_{\mathrm{BN}}) \leq 0 < \Phi(x_{\mathrm{p}})$,

$$x_{\mathrm{t}} = \frac{\Phi(x_{\mathrm{p}})}{E_{\mathrm{c}}} + x_{\mathrm{p}}, \tag{S17}$$

$$A = \frac{2}{3} q^{\frac{1}{2}} \left\{ \Phi_{\mathrm{B}}^{\frac{3}{2}} \frac{1}{E_{\mathrm{a}}} - \Phi(x_{\mathrm{n}})^{\frac{3}{2}} \left( \frac{1}{E_{\mathrm{a}}} - \frac{1}{E_{\mathrm{m}}} \right) - \Phi(x_{\mathrm{n}})^{\frac{3}{2}} \left( \frac{1}{E_{\mathrm{m}}} - \frac{1}{E_{\mathrm{c}}} \right) \right\}, \tag{S18}$$

$$B = 2q^{-\frac{1}{2}} \left\{ \Phi_{\mathrm{B}}^{\frac{1}{2}} \frac{1}{E_{\mathrm{a}}} - \Phi(x_{\mathrm{n}})^{\frac{1}{2}} \left( \frac{1}{E_{\mathrm{a}}} - \frac{1}{E_{\mathrm{m}}} \right) - \Phi(x_{\mathrm{n}})^{\frac{1}{2}} \left( \frac{1}{E_{\mathrm{m}}} - \frac{1}{E_{\mathrm{c}}} \right) \right\}. \tag{S19}$$

(Case 4) For $0 < \Phi(x_{\mathrm{BN}})$,

$$x_{\mathrm{t}} = T_{\mathrm{BN}}, \tag{S20}$$

$$A = \frac{2}{3} q^{\frac{1}{2}} \left\{ \Phi_{\mathrm{B}}^{\frac{3}{2}} \frac{1}{E_{\mathrm{a}}} - \Phi(x_{\mathrm{n}})^{\frac{3}{2}} \left( \frac{1}{E_{\mathrm{a}}} - \frac{1}{E_{\mathrm{m}}} \right) - \Phi(x_{\mathrm{n}})^{\frac{3}{2}} \left( \frac{1}{E_{\mathrm{m}}} - \frac{1}{E_{\mathrm{c}}} \right) - \Phi(x_{\mathrm{BN}})^{\frac{3}{2}} \left( \frac{1}{E_{\mathrm{c}}} \right) \right\}, \tag{S21}$$

$$B = 2q^{-\frac{1}{2}} \left\{ \Phi_{\mathrm{B}}^{\frac{1}{2}} \frac{1}{E_{\mathrm{a}}} - \Phi(x_{\mathrm{n}})^{\frac{1}{2}} \left( \frac{1}{E_{\mathrm{a}}} - \frac{1}{E_{\mathrm{m}}} \right) - \Phi(x_{\mathrm{n}})^{\frac{1}{2}} \left( \frac{1}{E_{\mathrm{m}}} - \frac{1}{E_{\mathrm{c}}} \right) - \Phi(x_{\mathrm{BN}})^{\frac{1}{2}} \left( \frac{1}{E_{\mathrm{c}}} \right) \right\}. \tag{S22}$$

These forms are reduced to the standard F-N tunneling equation by setting $E_{\mathrm{c}} = E_{\mathrm{m}} = E_{\mathrm{a}}$. In addition, the derived equations for case 2 correspond with those in reference 34. Furthermore, $A$ for all cases has been given in a form similar to that in reference 24, although the correction of the field by $B$ is not considered in the reference.

**Supplemental Material Note 2: List of notations.**

| | |
|---|---|
| $t$ | time [s] |
| $q$ | electron charge [C] |
| $h$ | Planck's constant [Js] |
| $m$ | electron mass [kg] |
| $m^*$ | electron mass in $h$-BN [kg] |
| $\varepsilon_{BN}$ | permittivity of $h$-BN [F/m] |
| $T_{BN}$ | thickness of $h$-BN [m] |
| $n_t(t)$ | density of trapped electron [1/m$^2$] |
| $p_t(t)$ | density of trapped hole [1/m$^2$] |
| $N_t$ | total neutral trap density for hole [1/m$^2$] |
| $n_G(t)$ | density of the total generated electron from $t = 0$ [1/m$^2$] |
| $n_R(t)$ | density of the total recombinated electron from $t = 0$ [1/m$^2$] |
| $p_R(t)$ | density of the total recombinated hole from $t = 0$ [1/m$^2$] |
| $\sigma_t$ | trap capture cross section area for hole [m$^2$] |
| $\sigma_R$ | trap capture cross section area for recombination [m$^2$] |
| $x$ | distance from the anode interface [m] |
| $x_n$ | centroid of $n_t$ [m] |
| $x_p$ | centroid of $p_t$ [m] |
| $x_{BN}$ | thickness of $h$-BN [m] |
| $x_t(t)$ | tunneling distance [m] |
| $\Phi(x, t)$ | potential based on the Fermi energy of anode [eV] |
| $\Phi_B$ | barrier height for hole at anode [eV] |

$E(x, t)$ electrical field defined as $-\mathrm{d}\Phi(x, t)/\mathrm{d}x$ [V/m]

$E_{\mathrm{BN}}$ average of field across the *h*-BN defined by $-\Phi(x_{\mathrm{BN}}, t) / x_{\mathrm{BN}} = -V_{\mathrm{BN}}/T_{\mathrm{BN}}$ [V/m]

$E_{ave}(t)$ average of field between $x_t$ and $x_{\mathrm{BN}}$ defined by $-(\Phi(x_{\mathrm{BN}}, t) - \Phi(x_{\mathrm{t}}, t)) / (x_{\mathrm{BN}} - x_{\mathrm{t}})$ [V/m]

$E_{\mathrm{c}}(t)$ cathode field [V/m]

$E_{\mathrm{m}}(t)$ field between $x_{\mathrm{p}}$ and $x_{\mathrm{n}}$ [V/m]

$E_{\mathrm{a}}(t)$ anode field [V/m]

$V_{\mathrm{BN}}$ potential drop across the *h*-BN (applied voltage) [V]

$\Delta V^{+}$ difference for the positive voltage at certain current in the *I-V* test [V]

$\Delta V^{-}$ difference for the negative voltage at certain current in the *I-V* test [V]

$j(t)$ current density [C/(sm$^2$)]

$\Delta j$ difference between measured current and the theoretical F-N tunneling current [C/(sm$^2$)]

$A(\mathrm{t})$ value which depends on a potential barrier for the F-N tunneling current equation [$\mathrm{C}^{1/2}\mathrm{V}^{1/2}\mathrm{m}$]

$B(\mathrm{t})$ value which depends on a potential barrier for the F-N tunneling current equation [$\mathrm{C}^{-1/2}\mathrm{V}^{-1/2}\mathrm{m}$]

$C_1$ material constant for the F-N tunneling current equation [$\mathrm{C}^2$/(Js)]

$C_2$ material constant for the F-N tunneling current equation [$\mathrm{kg}^{1/2}$/(Js)]

$\alpha(x, t)$ impact ionization coefficient [1/m]

$\alpha_0$ constant for impact ionization coefficient [1/m]

$H$ constant for impact ionization coefficient [V/m]

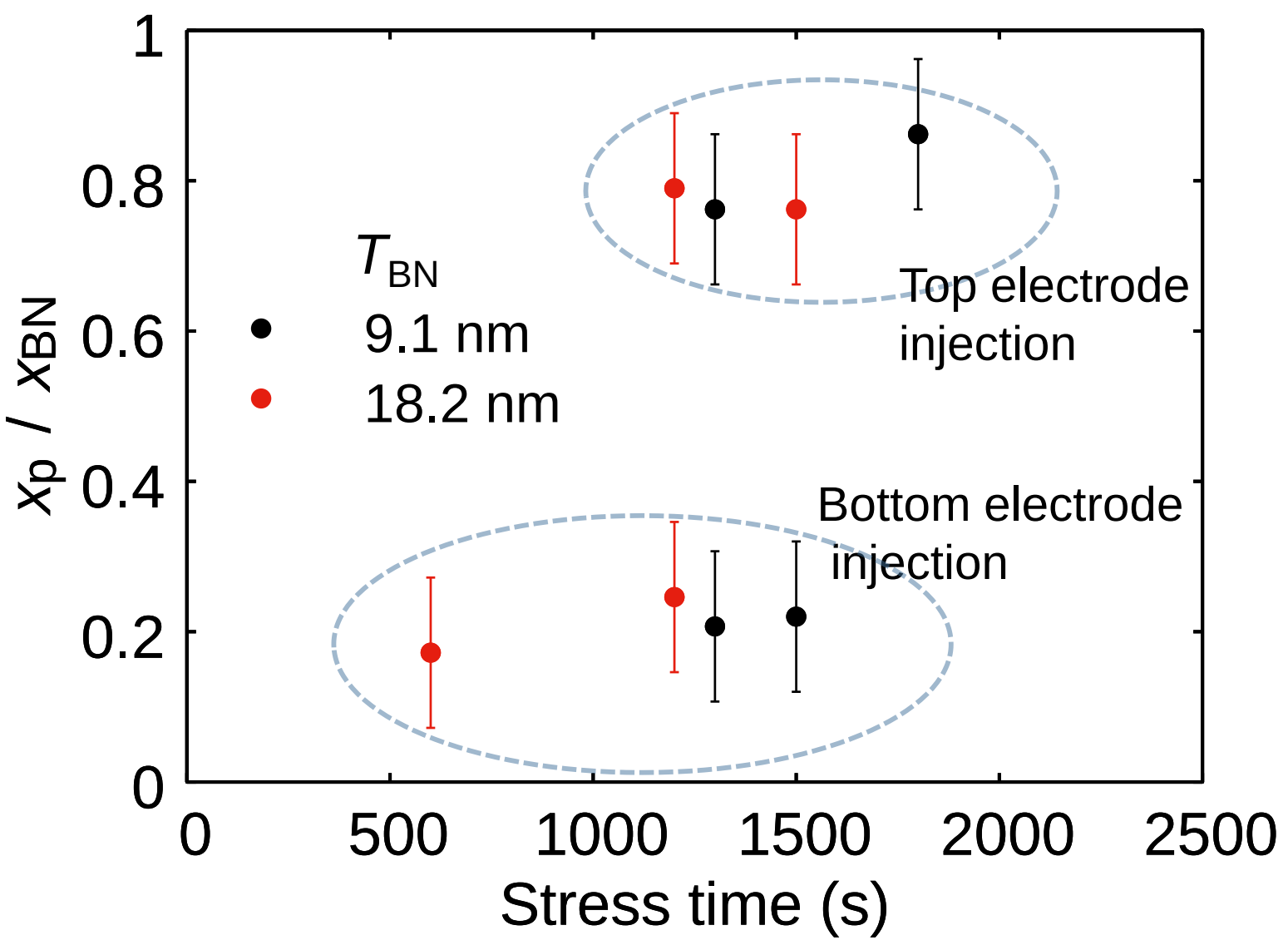

**FIG. S1:** Normalized $x_p$ as a function of the stress time for constant current at 5.0 ×10$^{-6}$ A/m$^2$. $x_p$ is normalized to the thickness of *h*-BN for comparison. $x_p/x_{BN}$ is independent of the stress time and the thickness of *h*-BN. When the sample was subjected to negative voltage stress (bottom electrode injection), $x_p/x_{BN}$ is approximately 0.2. In the case of the top electrode injection, it becomes 0.8, corresponding to 0.2 from the top electrode interface. These results suggest that electrode structure is ideally symmetrical with two graphite electrodes. Note that the constant-current measurement at ±1.0 ×10$^{-5}$ A/cm$^2$ for 1 s was adapted instead of the *I*-*V* measurement in the test for more proper estimation for $x_p$, which minimizes the electron trap during the measurement to obtain $\Delta V$.

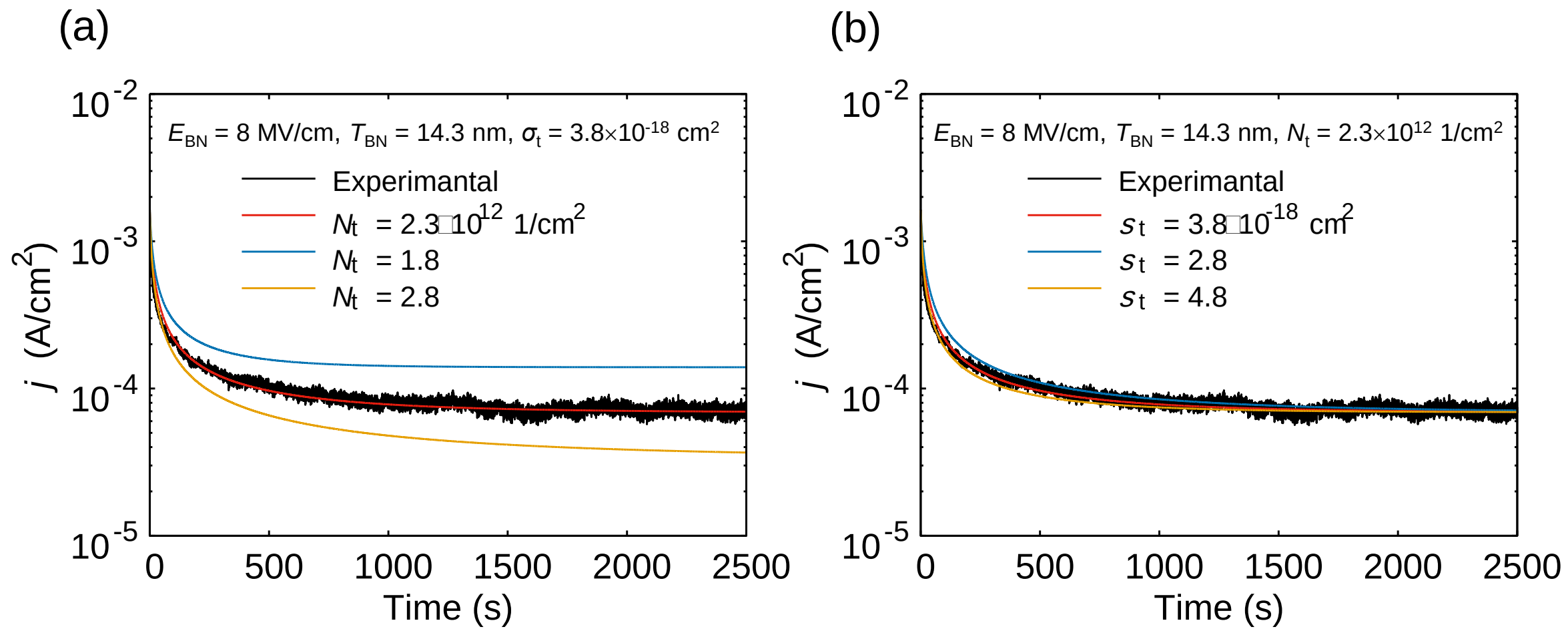

**FIG. S2**: Fitting of experimental data by the electron trap model, in which (a) $N_t$ and (b) $\sigma_t$ are used as fitting parameters.

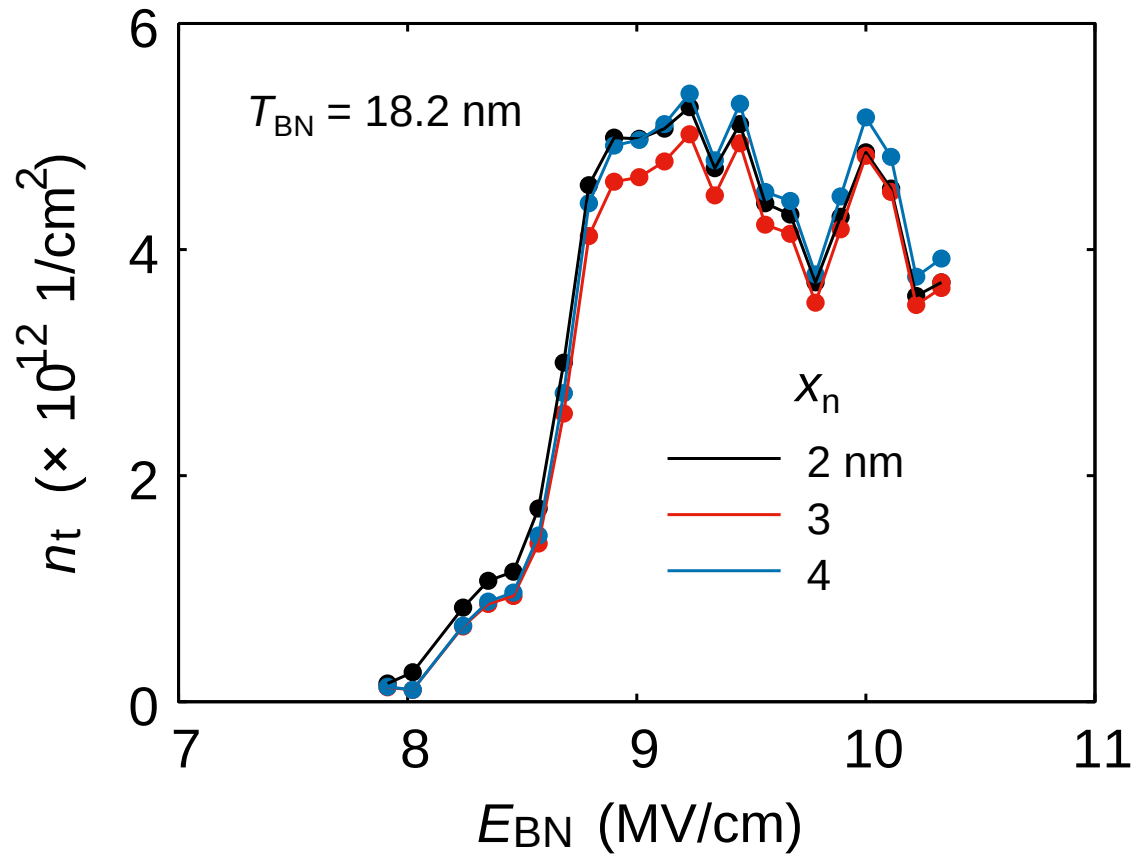

**FIG. S3**: Sensitivity of $n_t$ on $x_n$. $n_t$ is calculated for different values of $x_n$.

| Materials | $\alpha_0$ (1/cm) | *H* (MV/cm) | Reference |
|---|---|---|---|
| Ge | $6.39\times10^6$ | 1.27 | [49] |
| Si | $1.58\times10^6$ | 2.03 | [50] |
| InP | $2.03\times10^6$ | 2.34 | [52] |
| GaAs | $1.5\times10^6$ | 1.57 | [51] |
| 4H-SiC (//c) | $3.41\times10^8$ | 25 | [53] |
| 4H-SiC ($\perp$c) | $2.96\times10^7$ | 16 | [53] |
| GaN | $2.04\times10^6$ | 9.14 | [54] |
| Diamond | $3.3\times10^6$ | 14.2 | [50] |
| *h*-BN | $1.4\times10^{13}$ | 290 | This work |
| $SiO_2$ | $3.3\times10^6$ | 78 | [29] |

**FIG. S4**: The impact ionization coefficient for various material. All values are obtained for the impact ionization coefficient for hole except for $SiO_2$.